\documentclass{llncs}
\usepackage{times}
\usepackage{amssymb}
\usepackage{amstext}
\usepackage{amsmath}
\usepackage{color}
\usepackage{eucal}
\usepackage{graphics}
\usepackage{graphicx}
\usepackage[T1]{fontenc}
\usepackage{ae,aecompl}

\begin{document}
\title{Designing Electronic Markets for Defeasible-based Contractual Agents}

\author{Adrian Groza}
\institute{Technical University of Cluj-Napoca\\
Department of Computer Science\\
Baritiu 28, RO-3400 Cluj-Napoca, Romania\\
\email{Adrian.Groza@cs.utcluj.ro}\\
}

\maketitle
\thispagestyle{empty}

\begin{abstract}
The design of punishment policies applied to
specific domains linking agents actions to material penalties is
an open research issue.
The proposed framework applies principles of contract law to set
penalties: expectation damages, opportunity cost, reliance
damages, and party design remedies.  
In order to decide which remedy provides maximum welfare within an electronic market, a simulation environment called DEMCA (Designing Electronic Markets for Contractual Agents) was developed. 
Knowledge representation and the reasoning capabilities of the agents are based on an extended version of temporal defeasible logic. 

Keywords: contractual agents, legal remedies, defeasible logic.

\end{abstract}


\section{INTRODUCTION}
\label{sec:introduction}

According to~\cite{Vold02}, there are five different philosophies
of punishment from which all punishment policies can be derived:
deterrence, retribution, incapacitation, rehabilitation and
restoration. 
Retribution considers that the remedy should be as severe as the wrongful act, making this doctrine most suitable for multi-agent systems. 
Consequently, we drive our attention toward four legal doctrines from contract law that aim to equal the victim's harm: expectation damages, opportunity costs, reliance damages, and party designed remedies. 
The output of this research formalizes the above remedies for multi-agent systems within a framework for
computing penalties for B2B disputes. 

The remedies imposed by law affect the agents' behaviour~\cite{Cooter04}: 
(1) searching for trading partners; 
(2) negotiating exchanges; 
(3) keeping or breaking commitments; 
(5) taking precaution against breach causing events; 
(6) acting based on reliance on promises;
(7) acting to mitigate damages caused by broken commitments; 
(8) settle disputes  caused by broken promises.
In the supply chain context a contract breach can propagate over
the entire chain. 
The damages imposed by legal institutions can
positively influence breach propagation. 
Usually, a contract breach appears when some perturbation  arises on the
market\footnote{For instance, the market price of a raw material
could rise so much that, for the agent who had planed to achieve
it in order to produce an item, is more efficient to breach the
contract with its buyer.}. 
The question is which of the above remedies is adequate for
an efficient functionality of the electronic market. Since normative
reasoning is defeasible by nature, we developed a framework which provide defeasible reasoning capabilities for agents to act in case of breach.  

The problem in hand is approached by: 
(i) formalising contractual clauses and contract law remedies for multi-agent systems; 
(ii) providing a system for designing experiments in order to decide which legal doctrine suits an electronic market;
iii) providing defeasible logic-based mechanisms for enhancing reasoning capabilities of the agents about penalties.
The paper is organized as follows: 
The next section introduces contracts within the task dependency network model.
Section~\ref{sec:remedies} formalises four types of
remedies for multi-agent systems according to contract law.
In section~\ref{sec:analyses} the functions used by the market for
penalties are implemented. 
Section~\ref{sec:demca} describes the implemented system. 
Section~\ref{sec:related} details related work and section~\ref{sec:conclusions} concludes the paper.

\section{PROBLEM SPECIFICATION}
\label{sec:design}

\subsection{Commitment Dependency Network}

The task dependency network model~\cite{Walsh03}, used in the
analysis of the supply chain, was adapted~\cite{Letia05IAT} as follows:
commitment dependency network is a directed, acyclic graph,
(V,E), with vertices $V = G \cup A$, where:
$G$ = the set of goods,
$A = S \cup P  \cup C$  the set of agents,
$S$ = the set of suppliers,
$P$ = the set of producers,
$C$ = the set of consumers,
and a set of edges E (commitments) connecting agents with their input and output goods.
\begin{figure}
    \begin{center}
        \includegraphics[width=0.75\textwidth]{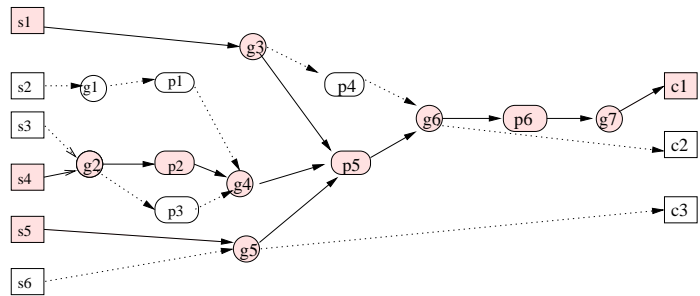}
    \end{center}
    \caption{Task dependency network: goods are indicated by circles,
suppliers and consumers are represented by boxes, while producers by curved boxes.}
    \label{fig:task}
\end{figure}
With each agent $a \in A$ we associate an input set $I_a$ and an output set $O_a$:
$I_a=\{g \in G |\prec g,a \succ \in E\}$ and $O_a=\{g \in G |\prec a,g \succ \in E\}$.
Agent $a$ is a supplier if $I_a=\emptyset$, a consumer if $O_a=\emptyset$, and a producer in all other cases. Without any generalization lost, we consider that a consumer $c \in C$ needs a single item ($|I_c|=1$) and every supplier $s \in S$ or producer $p \in P$ build one single item ($|O_s|=1$ and $|O_p|=1$)

An agent must have a contract for all of its input goods in order to produce its output,
named \emph{presumable}
\footnote{Note that when someone breaches a contract with a presumable agent it has to pay more damages.} and denoted by $\hat{p}$.
If we note $n_p=|I_p|$, the agent has to sign $n_p+1$
contracts in order to be a member in the supply chain.
For each input good $g_k \in I_p$ the agent $p$ bids its own item valuation $v_p^k$.
The auction for the good $g_k$ sets the transaction price at $p_k$.
The agent's investments are $I_p=\sum_{k=1}^{n_p}{p_k}$ where $k$ are the winning input goods.
We note by $I_p^g$ the agent's investments but without considering the investments made for the current good $g$.
Similarly, we note all bids values submitted by the agent $p$ as $V_p=\sum_{k=1}^{n_p}{v_p^k}$ and this value without considering the bid for good $g$ as $V_p^g$.
For the output good, the agent $p$ signs a contract at reliance price $R_p$.
We consider that there are no production costs and
when perturbation or unexpected events occur,
agents need  protocols for repairing or reforming the supply chain.

\subsection{Contracts}
By extending social commitments~\cite{Pasquier05lnai}, we define a set of contractual patterns used to declaratively specify the contracts signed between parties. 
The classical definition of a conditional commitment states that a commitment is a promise from 
a debtor $x$ to a creditor $y$ to bring about a particular sentence $p$
under a condition $q$.
Starting from this definition we formalise a set of contractual clauses inspired from contract law:

In a \textit{Gratuitous Promise} the debtor $x$
promises the creditor $y$ to bring about $p$ until
$t_{maturity}$ without requesting anything: $C^0_1(x,y,1,p_1, t_{maturity})$. 
A \textit{Unilateral Contract} involves an exchange of the offerer's promise $p$ for the oferee's act $q$, where the debtor $x$ promises the creditor $y$ to bring about $p$ until $t_{maturity}$
if condition $q$ holds at time $t_{issue}$.
For instance, in the contract
$C=\prec a_s, a_b, g_i, P_c, t_{issue}, t_{maturity}\succ$, 
$a_s$ represents the seller agent, $a_b$ the buyer agent, $g_i$ the good or the transaction subject, $P_c$ the contract price, $t_{issue}$ is the time when the offer is accepted and $t_{maturity}$ is the time when the transaction occurs.
In a \textit{Bilateral Contract} both sides make promises, 
the debtor $x$ promises the creditor $y$ to bring
about $p$ if the creditor $y$ promises $x$ to bring about $p_1$, formalised as $C(x,y,C(y,x,1,p_1),p)$. 

During its life cycle, a commitment may be in one of the following states: 
\textit{open offer, active, released, breached, fulfilled, canceled, or failed}, which are also useful to be considered from a legal perspective. 
The state transition between \textit{open offer} and \textit{active} contract depends on each type of commitment: the acceptance of a gratuitous commitment means reliance and acting upon it, the acceptance of a unilateral contract means the execution of the required task, the acceptance of a bilateral contract means the creation of the required promise.


\section{REMEDIES}
\label{sec:remedies}

The remedies described in this section try to equal the victim's harm. 
In the first three
cases\footnote{Expectation damages, reliance damages and
opportunity cost are analyzed from an economical point of view
in~\cite{Cooter04,Friedman00}.}, the system estimates the harm
according to current market conditions, while in the last case,
the agents themselves compute the damages and generate their own
penalties.

\paragraph{Expectation Damages.} The courts reward damages that place the victim of breach in the position he or she would have been in if the other party had performed the contract~\cite{Cooter04}.
Therefore, in an ideal situation, the expectation damages does not affect the potential victims whether the contract is performed or breached.
Ideal expectation damages remain constant when the promisee relies on the performance of the contract more than it is optimal.

\paragraph{Reliance Damages.} Reliance increases the loss resulting from the breach of the contract. 
Reliance damages put the victim in the same position after the breach as if he had not signed a contract with the
promisor or anyone else~\cite{Cooter04}. In an ideal situation,
the reliance damages do not affect the potential victims if
the contract is breached or there was no initial contract. No
contract provides a baseline for computing the injury. Using this
baseline, the courts reward damages that place the victims in the position that they would have been, if they had
never contracted with another agent. 
Reliance damages represent
the difference between profit if there is no contract and the
current profit.

\paragraph{Opportunity Cost.} Signing a contract often entails the loss of an opportunity to make an alternative.
The lost opportunity provides a baseline for computing the damage.
Using this baseline, the courts reward damages that place victims of breach in the position that they would have been if they had signed the contract that would have been the best alternative to the one that was breached~\cite{Cooter04}.
In the ideal situation, the opportunity cost damages does not affect the potential victims whether the contract is breached or the best alternative contract is performed\footnote{Opportunity cost and expectation damages approach equality as markets approach perfect competition.}.
If breach causes the injured party to purchase a substitute item, the opportunity cost formula equals the difference between the best alternative contract price available at the time of contracting and the price of the substitute item obtained after the breach.

\paragraph{Party-Designed Remedies.} The contract might stipulate a sum of money that the breaker will
pay to the party without guilt. These "leveled commitment
contracts"~\cite{Sandholm01} allow self-interested agents to face
the events that unfolded since the contract started. A rational
person damages others whenever the benefit is large enough to pay
an ideal compensation and have some profit, as required to
increase efficiency. Game theoretic analysis has shown that
leveled committed contracts increase the Pareto efficiency. One contract may charge a high price and offer to pay
high damages if the seller fails to deliver the goods, while
another contract may charge a low price and offer to pay low
damages, the types of contracts separating the set of buyers and
allowing "price discrimination."



\section{CASE ANALYSIS}
\label{sec:analyses}

The conclusions from the last sections are:
(i) The amount of expectation damages must place the victim
in the same position as if the actual contract had been performed; 
(ii) The amount of reliance damages must place the victim in the
same position as if no contract had been signed; (iii) The amount
of opportunity-cost damages must place the victim in the same
position as if the best alternative contract had been performed;
(iv) Party designed remedies specify themselves the amount of
damages in case of a breach.

\subsection{No substitute}

\subsubsection{Supplier-Consumer}

\begin{figure}
    \begin{center}
        \includegraphics[width=0.60\textwidth]{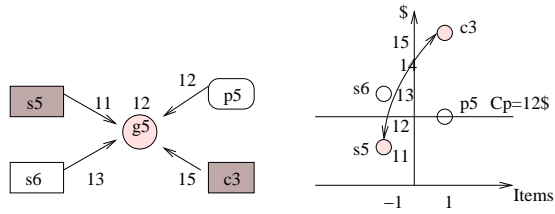}
    \end{center}
    \caption{Supplier-Consumer contract}
    \label{fig:supplier-consumer}
\end{figure}

\paragraph{The consumer breaches the contract.}
In fig.~\ref{fig:supplier-consumer}a) the suppliers $s_5$ and
$s_6$ want to sell good $g_5$ at price 11 and respectively 13,
while the agents $p_5$, and $c_3$ try to buy it at prices 12 and
15. According to (M+1)st price protocol the transaction price is
$P_c=12\$$\footnote{The goods are transacted using the (M+1)st price auction
protocol, which has the property to balance the
offer and the demand at each level in the supply chain (otherwise
the supply demand equilibrium cannot be achieved globally). 
It provides a uniform price mechanism: all contracts determined by a
particular clearing are signed at the same price. 
}. The auction clears at every round. In
fig.~\ref{fig:supplier-consumer}b) a single unilateral contract is signed:
$C_{g_5}^1=\prec s_5,c_3,g_5,12,t_{issue},t_{maturity}\succ$. 
Consider $c_3$ breaches the contract. In this case, the
remedies will be:
i) \emph{Expectation damages}:
if the agent $c_3$ performs, the $s_5$'s estimated profit is the difference
between the contract price $P_c=12$ and its own
valuation\footnote{(M+1)st price auction has the following property: the dominant
strategy for each agent is to reveal its real valuation.}
$v_{a_6}^{g_5}=11$ (victim valuation).
The remedies compensate this value: $D_e=P_c-v_a^g$.
; ii) \emph{Opportunity damages}: first, the auctioneer has to compute
the opportunity cost $P_o$, which is the transaction cost in case
the breacher was absent from the auction. In
fig.~\ref{fig:supplier-consumer}, if agent $c_3$ is not present
$P_o=11$. The $s_5$'s bid is one who wins. The contract would be
$C_{g_5}^1=\prec s_5,p_5,g_5,11,t_{issue},t_{maturity}\succ$ and
the agent's profit would be $P_o-v_a^g$. But, when there is no
contract for agent $s_5$, his profit would be null. The
opportunity damages should reflect this. We define opportunity
cost damage $D_o$ which is received by the agent $a$ as:
$D_o=max(P_o-v_a^g,0)$; 
iii) \emph{Reliance damages}: if the victim does not have any input
good, the supplier's investments in performing are null: $D_r=0$; 
iv) \emph{Party-designed remedies}: the remedies may be a fraction from the contract price ($D_p=\alpha\cdot P_c$), a fraction from the expected profit ($D_p=\alpha\cdot D_e$), or constant ($D_p=C)$.

\paragraph{The supplier breaches the contract.}

Consider $s_5$ breaches contract $C_{g_5}^1$: 
i) \emph{Expectation damages}: $D_e=v_a^g-P_c$; 
ii) \emph{Opportunity damages}: if the breacher had not bid and the victim had signed a contract at the opportunity price $P_o$, than it's profit would have been $v_a^g-P_o$. 
If the victim has no contract when the breacher is not bidding, it receives no damages.
Hence, $D_o=max(v_a^g-P_o,0)$. 
In the depicted case, if the agent $s_5$ had not existed, $c_3$ would have signed a contract with $s_6$ for an opportunity cost $P_o=12$. Therefore, $D_o=3$; 
and iii) \emph{Reliance damages}: because the client does not produce any output goods,
it's reliance is null: $D_r=0$.

\subsubsection{Supplier-Producer}
\begin{figure}
    \begin{center}
        \includegraphics[width=0.65\textwidth]{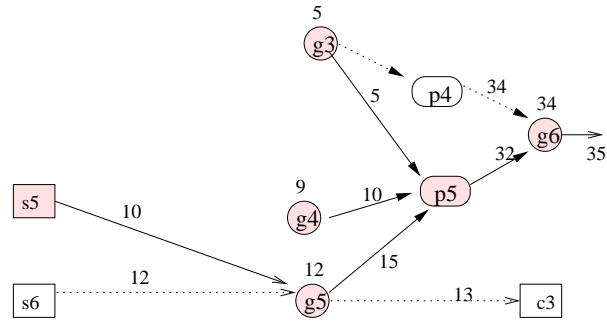}
    \end{center}
    \caption{Supplier-Producer contract}
    \label{fig:supplier_producer}
\end{figure}

\paragraph{The supplier breaches the contract.}
Consider the contract from fig.~\ref{fig:supplier_producer} $C=\prec s_5, p_5,g_5,12,t_{issue},t_{maturity}\succ$:  
i) \emph{Expectation damages:}
Observe that the victim is a presumable agent because it has contracts for all its input goods.
Its investments are $I_p=5+9+12=26$ and $I_p^{g_5}=9+5=14$.
The producer $p_5$ has also a contract for its output item, so $R_{p_5}=34$.
Its profit is $R_{p_5}-I_p=8$.
When bad contracts have been signed this value can be negative, therefore no damages are imposed.
Otherwise, expectation damages equals the difference between its bid and the contract price:
\label{subsec:supplier-producer}

\begin{minipage}[b]{0.4\linewidth}
$$D_e=
\begin{cases}
max(R_p-I_p,0), & \text{$\hat{p}$, $\exists{R_p}$}\\
v_p^g-P_c,  &\text{otherwise}
\end{cases}
$$
\end{minipage}
\hspace{0.1cm}
\begin{minipage}[b]{0.6\linewidth}
$$D_r=
\begin{cases}
V_p^g-I_p^g+R_p-v_p^{g_o},  & \text{$\hat{p}$, $\exists{R_p}$}\\
V_p^g-I_p^{g_k},    &\text{otherwise}
\end{cases}
$$
\end{minipage}

$$D_o=
\begin{cases}
max(R_p-I_p^g-P_o,0),   & \text{$\hat{p}$, $\exists{R_p}$, $\exists{P_o}$}\\
v_p^g-P_o,  &\text{$\neg \hat{p}$, $\exists{R_p}$, $\exists{P_o}$}\\
0,  &\text{$\neg{\exists{P_o}}$}
\end{cases}
$$

Recall $\hat{p}$ means that agent $p$ is presumable, $g_o$ is the output good of the agent $p$ and $I_p^{g_k}$
represents all $k$ contracts signed for input goods, where
$k<n_p$. In the depicted case $p_5$ is presumable and there is a
contract with a buyer. Therefore, it has to receive, as a victim,
the next reliance damages
$D_r=V_{p_5}^{g_5}-I_{p_5}^{g_5}+R_{p_5}-v_{p_5}^{g_6}=(10+5)-(9+5)+34-32=3$.
In case of \emph{Opportunity costs} damages, one seller less implies $P_o\geq P_c$. 

In some cases damages can be higher than the contract value itself
($D_r \geq P_c$). According to current practice in law, these
damages are the right ones if the victim gives a previous
notification about the risks faced by the potential breacher. This
is a clear situation when information propagation improves the supply
chain performance. In the light of the above facts, their reliance
damages should remain the mentioned ones if the victim has
notified its partner, but should be maxim $P_c$ otherwise. Hence,
we define $D'_r$ equals $D_r$ when the breacher receives a notice, 
and $D'_r =min(D_r,P_c)$, otherwise.

\subsubsection{Producer-Consumer}
\begin{figure}
    \begin{center}
        \includegraphics[width=0.65\textwidth]{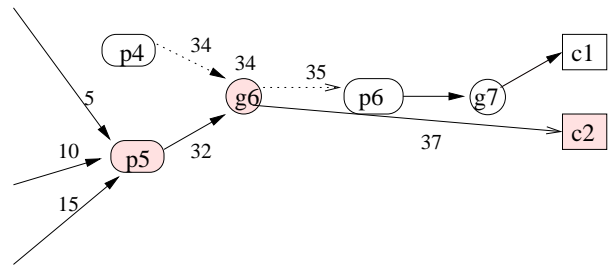}
    \end{center}
    \caption{Producer-Consumer contract}
    \label{fig:producer_consumer}
\end{figure}

\paragraph{The consumer breaches the contract}
Consider the unilateral contract $C=\prec p_5, c_2,g_6,34,t_{issue},t_{maturity}\succ$ from fig.~\ref{fig:producer_consumer}, where $c_2$ breaches: 
i) In the case of \emph{Expectation damages}, $p_5$ is presumable and $D_e=34-(12+9+5)=8$.
Suppose the agent $p_5$ does not have any contract for one input good.
Therefore, it is not presumable and it will receive $D_e=34-32$; 
ii) \emph{Reliance damages}:
$D_r=V_p-I_p$; and 
iii) \emph{Opportunity cost}: one buyer less implies $P_o\leq P_c$:

\begin{minipage}[b]{0.45\linewidth}
$$D_e=
\begin{cases}
max(P_c-I_p^g,0),   & \text{$\hat{p}$}\\
P_c-v_p^g,  &\text{otherwise}
\end{cases}
$$
\end{minipage}
\hspace{0.2cm}
\begin{minipage}[b]{0.55\linewidth}
$$D_o=
\begin{cases}
max(P_o-I_p^g,0),   & \text{$\hat{p}$, $\exists{P_o}$}\\
P_o-v_p^g,  &\text{$\neg \hat{p}$, $\exists{P_o}$}\\
0,  &\text{$\exists \overline{P_o}$)}
\end{cases}
$$
\end{minipage}

\subsection{Substitute}

The common law requires the promisee to mitigate damages.
Specifically, the promisee must take reasonable actions to reduce losses occurred by the promisor's breach.
The market can force the victim to find substitute items, in this case the imposed damages reflect only the difference between original contract and substitute contract.
With a substitute contract, the victim signs for the identical item, with the same deadline or $t_{maturity}$, but at a different price.
Let $P_s$ be the value of the substitute contract.
For the general case \emph{Producer-Producer}, when the buyer breaches the contract, the equations become:

\begin{minipage}[b]{0.45\linewidth}
$$D_e=
\begin{cases}
max(P_c-I_p^g,0),   & \text{$\hat{p}$, $\neg  \exists P_s$ }\\
P_c-v_p^g,  &\text{$\hat{p}$, $\neg  \exists P_s$}\\
max(P_c-P_s), &\text{$\exists P_s$}
\end{cases}
$$
\end{minipage}
\hspace{0.2cm}
\begin{minipage}[b]{0.55\linewidth}
$$D_r=
\begin{cases}
V_p-I_p,    & \text{$\neg  \exists P_s$}\\
max(P_c-P_s,0), &\text{$\exists P_s$}
\end{cases}
$$
\end{minipage}

$$D_o=
\begin{cases}
max(P_o-I_p^g,0),   & \text{$\hat{p}$, $\exists{P_o}$, $\neg  \exists P_s$}\\
max(P_o-v_p^g,0),   &\text{$\neg \hat{p}$,$\exists{P_o}$, $\neg  \exists P_s$ }\\
max(P_o-P_s,0), &\text{$\exists P_s$}
\end{cases}
$$

\section{DEMCA: A SIMULATION ENVIRONMENT FOR CONTRACT BREACH EXPERIMENTS}
\label{sec:demca}

\subsection{System Architecture}
\begin{figure}
    \begin{center}
        \includegraphics[width=0.65\textwidth]{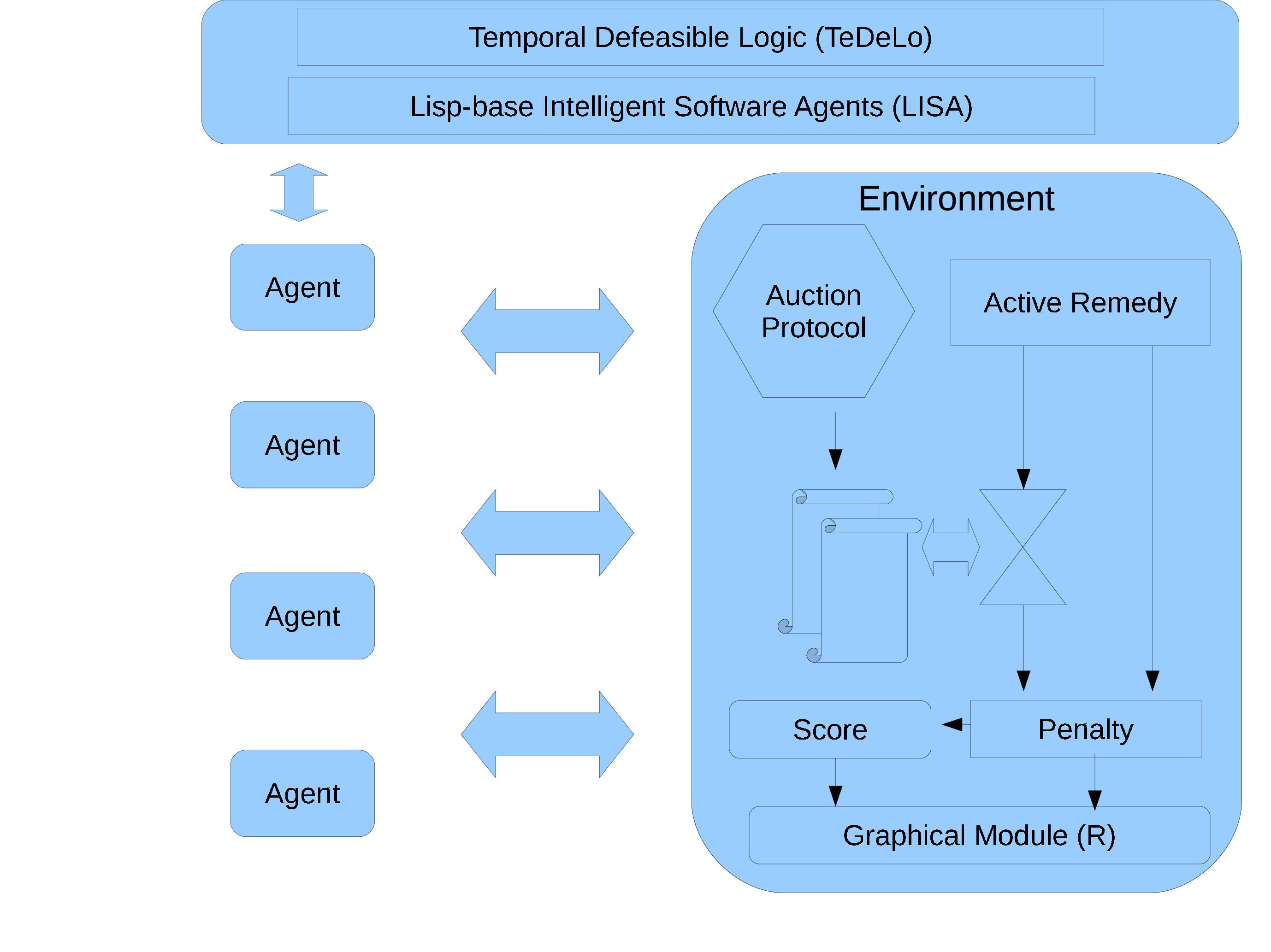}
    \end{center}
    \caption{System architecture.}
    \label{fig:demca}
\end{figure}
Designing Electronic Markets for Contractual Agents (DEMCA) tool\footnote{Available at http://cs-gw.utcluj.ro/$\sim$adrian/odr.html} consists of the next three components (figure~\ref{fig:demca}):

\subsubsection{Agents.}
For defining a contractual agent in DEMCA one has to instantiate a generic agent with the following structure (figure~\ref{fig:demca_agent}). 
The \textit{type} of the agent identifies its role in the supply chain: supplier, producer, or consumer.
The agent's strategy is defined by the function \textit{program}, which points to a set of temporalised defeasible rules coded in the TeDeLo (Temporal Defeasible Logic) framework. 
Examples of strategies include: reactive/proactive agents, breaching often/seldom agents, information propagation/hiding agents. 
The agent has a list o percepts consisting in public information provided by the market: the current open auctions, the existing bids, the agents, the public contracts. 
Each agent has attributes keeping \textit{stored items}, \textit{active contracts}, or \textit{the amount of money} available at the beginning of each experiment round (figure~\ref{fig:demca_agent}).
Other information that can be provided is: the goods which the agent is interested in achieving, the good that the agent produces, the own valuation of these items. 
The public or private character if these information depends on the current simulation.
\begin{figure}[t]
\footnotesize
\begin{verbatim} 
(defun generic_agent (name &optional (program (program name)) stored_items 
                      active_contracts input_goods output_goods valuation_in 
                      valuation_out money type))
(MAKE-agent :name name :type type :program program
            :legal-actions '(bid breach) :stored_items stored_items
            :active_contracts active_contracts :score money
            :input_goods input_goods :output_goods output_goods
            :valuation_in valuation_in :valuation_out valuation_out))
\end{verbatim}
  \caption{Agent's structure in DEMCA.}
\label{fig:demca_agent}
\end{figure}

\subsubsection{Environment.}
The simulation environment introduces auctions for each item within the market, to which the agents manifest their buying or selling interests.
The market is governed by a discrete time step. 
At the beginning of each round the agent perceives the public information, and after a reasoning process, it conveys one of the following actions: \textit{bid}, \textit{breach}, \textit{no\_action}.  
A contract might be signed in case the agent wins the M+1 price auction. 
If the agent has all the necessary preconditions, the environment simulates the contract execution. 
Otherwise, if the preconditions are not met at the time of maturity and no \textit{breach} action was executed, the environment automatically computes the amount of remedy according to the legal doctrine in force.

The auctions run in parallel, following the same M+1 price protocol. 
After the bids were sent, contracts are generated for the winning agents, the public information being broadcasted to all the agents within the task dependency network. 
The environment is responsible for computing the profit for each agent, profit which includes the possible penalties.
During DEMCA experiments five types of remedies can be used: expectation damages, opportunity costs, reliance 
damages, fixed penalty, and percentage from the contract value.
The output of the simulation is analysed by some scripts coded in $R$ language, responsible for generating graphics presenting the evolution of the agents' score during experiments. 

\subsubsection{Reasoning capabilities.}
The framework provides defeasible reasoning capabilities to the DEMCA agents to flexibly represent market strategies under time constraints and levels of certainty for handling preferences when deciding to breach or not a contract. 
By enhancing agents with defeasible reasoning, they are able to  reason with incomplete information, including confidential
contractual clauses. On the other hand, defeasible logic has already been proved to be suitable for legal reasoning~\cite{Hage03AIL}.

The expressivity of \textit{defeasible logic} is given by three types of rules:
\textit{Strict rules} are rules in the classical sense, that is
whenever the premises are indisputable, then so is the conclusion, while
\textit{defeasible rules} are rules that can be defeated by contrary evidence.
For "sending the goods means the goods were delivered",
if we know that the goods were sent then they reach the destination,
unless there is other, not inferior, rule suggesting the contrary.
\textit{Defeaters} are rules that cannot be used to draw any conclusions.
Their only use is to prevent some conclusions, as in
"if the customer is a regular one and he has a short delay for paying,
we might not ask for penalties".
This rule cannot be used to support a "not penalty" conclusion,
but it can prevent the derivation of the penalty conclusion.
Defeaters have a particular role in this context due to their capability to explicitly represent exception, or contract breaches in our case.

Defeasible derivations have an argumentation like structure:
firstly, we choose a supported rule having the conclusions $q$ we want to prove,
secondly we consider all the possible counterarguments against $q$, and
finally we rebut all the above counterarguments showing that,
either some of their premises do not hold, or the rule used for its derivation
is weaker than the rule supporting the initial conclusion $q$.

The TeDeLo (Temporal Defeasible Logic) module implements an extended version of temporal defeasible logic ETDL, based on LISA (Lisp based Intelligent Software Agents). 
Compared to the existing systems, the current implementation extends the defeasible logic
with: 
(1) temporal defeasible reasoning, 
(2) certainty factor instead of a partial order relation, 
(3) open world assumption applied to the rules, 
(4) dynamic rules, 
(5) dynamic priorities, 
(6) identifying of hidden rules, 
(7) pattern of rules for representing cyclic events or facts, 
(8) rollback activation to better handle exceptions.
The formalism was extended with temporal aspects in order to reason with contract deadlines. 
Dynamic rules and priorities allow agents to adjust their market strategies according to the changes in the environment. 

\subsection{Running scenarios}
\label{sec:scenarios}
\subsubsection{Basic Example}
In order to trace a simple scenario, two agents were added to the framework, by calling the function \textit{add\_agent}:

\begin{small}
\begin{verbatim}
(add_agent 'reactive_agent 's15 '() '() '()   '(g5) 5  11 100)
(add_agent 'reactive_agent 'c4  '() '() '(g5) '()   15 16 100)
\end{verbatim} 
\end{small}

Here, the supplier $s_{15}$ is interested in selling the item $g_5$ for a minimum price of $11$ and needs 5 units to produce it, while the client $c_4$ is interested in buying the item $g_5$ at a maximum price of $15$, its own valuation being $16$.  
At the beginning of the experiment both agents have 100 monetary units and no stored item.  
\begin{figure}[t]
\begin{footnotesize}
\begin{verbatim} 
Time step 1:
Agent S15, 100, NIL, NIL, (G5) perceives: 
((#S(GOOD :NAME G5
          :SELL_AGENTS (S15, 100, NIL, NIL, (G5))
          :BUY_AGENTS (C4, 100, NIL, (G5), NIL)
          :SELL_BIDS NIL
          :BUY_BIDS NIL
          :CONTRACT NIL)))
and does:
(((BID #S(GOOD :NAME G5
               :SELL_AGENTS (S15, 100, NIL, NIL, (G5))
               :BUY_AGENTS (C4, 100, NIL, (G5), NIL)
               :SELL_BIDS NIL
               :BUY_BIDS NIL
               :CONTRACT NIL))))

\end{verbatim}
\end{footnotesize}
  \caption{Trace of the agent's percepts and actions in DEMCA.}
\label{fig:trace}
\end{figure}

Figure~\ref{fig:trace} presents part of a DEMCA session. 
During the first round, the environment automatically starts an auction for the item $g_5$. 
The agent $s_{15}$ perceives this open auction, the list of agents which are interested in selling the item (only itself in this case), the list of agents interested in buying it (the consumer $c_4$), the current buying or selling offers, at the output in case the auction clears.
Consider, the agent $s_{15}$ has a reactive strategy, in which if it has an item for selling and no active contract for it, the agent bids its own valuation of the item. 
This is formalised as a defeasible theory as follows:

\begin{small}
\begin{verbatim}
(defeasible 'r1 0.8 'neg 'contract 1 1 'poz bid 1 3) 
(defeater 'r2 0.9 'poz 'win 1 1 'neg bid 2 3)
\end{verbatim} 
\end{small}
Here, the defeasible rule $r_1$ supports with a certainty factor of $0.8$ the consequent $bid$ for all time instances within the interval $[1,3]$ in case the agent has not signed a contract for the desired item (\textit{'neg contract}) during the first round.
In defeasible reasoning, the conclusion of a defeasible rule can be blocked by contrary evidence, in case it has less degree of support. 
In the depicted case, the defeater $r_2$ may rebuttaly block the derivation of the \textit{bid} consequent, by supporting the opposite conclusion \textit{'neg 'bid}. 
In case the agent won the auction at timestep 1 (\textit{'win 1 1}), the \textit{bid} action is no longer derived because the defeater $r_2$ provides a stronger degree of support. 
In DEMCA, one can dynamically adjust these parameters, in order to tune the strategy to the market conditions. 
By using \textit{patterns of rules} facility provided by the platform, the agent can specify cyclic activations of specific events. 

Browsing the percepts sensed at the beginning of each round, the agent can incorporate a part of them into his defeasible theory, using the function \textit{(fact 'neg 'contract 1.0 1 1)}, which stipulates that the agent does not have a contract at round 1. 
Because the information is considered sure, the certainty factor of this fact is set to 1.0. 
Having this fact asserted, the supplier $s_{15}$, according to rule $r_1$, will execute the \textit{bid} action (figure~\ref{fig:trace}).  

\subsubsection{Large scale experiments}

First, the framework can be used as a tool for automated online
dispute resolution (ODR). There are three situations:
(i) The market may have substantial authority, hence \emph{one
remedy is imposed to all agents}. In this case, the amount of
penalties can be automatically computed with this framework; (ii)
Consistent with party autonomy, \emph{the agents may settle on
different remedies at contracting time}. This approach increases
flexibility and efficiency, because the agents are the ones
who know what type of remedy better protects their interests;
(iii) \emph{All the above remedies influence the amount of
penalties}: in this approach the role of the framework is to
monitor the market and collect data such as: the expected profit,
the opportunity cost, the amount of investments made, if there is
a substitute at $t_{breach}$. All these information is used as
arguments when the dispute is arbitrated~\cite{Toulmin58} in an
architecture which combines rule based reasoning (laws) and case
based reasoning (precedent cases). 

Second, knowing the bids, the actual contracts, the amount of
potential remedies, and the available offers on the market, the
framework can identify situations in which for both agents is more
profitable to breach the contract when a fortunate or an
unfortunate contingency appears. It computes pairs of suggestions,
helping to increase total welfare towards Pareto frontier.

Third, as a simulation tool, the market designer may obtain
results regarding the following questions: what types of remedies
assure flexibility in the supply chain? or
how information sharing influences total revenues or can be used to
compute optimum reliance? 
In the developed prototype we are currently making experiments with different types of agents:
low-high reliance, breach often-seldom, sharing information-not sharing, risk seeking-averse (when they are risk averse, the penalties do not need to be so high to force breachers behave appropriately). 


\section{RELATED WORK}
\label{sec:related} 

The task dependency network model was
proposed~\cite{Walsh03} as an efficient market mechanism in
achieving supply chain coordination. The authors analyze protocols
for agents to reallocate tasks for which they have no acquired
rights. However, this approach is rather a timeless-riskless
economy. On real markets a firm seldom signs contracts with its
buyers and its suppliers simultaneously. Moreover, the breach of a
contract implies no penalties, which is an unrealistic assumption
in real world. In contrast, in the DEMCA framework auctions end independently, and we introduce penalties in case of contract
breaching.  

The role of sanctions in multi-agent systems~\cite{Chaib-draa04}
is the enforcement of a social control mechanism for the
satisfaction of commitments. We focus only on material sanctions
and we do not include social sanctions which affect trust,
credibility or reputation. Moreover, we have applied four types of
material remedies in a specific domain. 
In the same spirit of
computing penalties according to the level of harm produced, the
amount of remedies may depend on the time when the contract was
breached~\cite{Toledo01}.
Expectation damages, reliance damages, 
and opportunity costs have also been
studied~\cite{Craswell00,Cooter04,Friedman00} and how contracts influence the supply chain coordination or strategic breach appear in~\cite{Cachoon00,Sandholm01}.

Experiments regarding breach penalties were attacked from a game theory viewpoint in~\cite{Sandholm01}. 
The contract specifies decommitment penalties for both parties. 
The resulting leveled commitment contracts enable agents to sign agreements that would be inefficient in case of full commitment contracts. 
The efficiency increases according to the fitness of the penalty to market conditions.
Our hypothesis is that, even if the amount is not known apriori the penalties introduced in this paper can be proved efficiently because they perfectly reflect the environment conditions. 
Further experiments are required to validate this idea. 
There are two issues using levelled commitment contracts in real life scenarios.
On the one hand, a contract has penalties associated to the main contractual clauses. 
When a situation which has not been enclosed into the penalty clauses occurs, the system still has to provide a  mechanism of computing remedies.  
On the other hand, the agents are legally bind to a penalty if there is a normative theory beyond that penalty. 
Otherwise, the agent can invoke many legal doctrines in order to avoid a penalty. 
This is not the case in DEMCA framework.
It is based on such normative theory, and in case of unexpected exceptions one can apply legal doctrines in order to compute the fair amount of penalty. 

As commitments appear to be sometimes too restrictive (direct obligations) and sometimes too flexible, 
enriched agents with defeasible reasoning mechanism, the resulting \textit{defeasible commitments} are more flexible than usual obligations but also more constrained than permissions~\cite{Pasquier05lnai}.
Dr-contract~\cite{Governatori05} also provides defeasible reasoning mechanism for representing business rules.
By introducing temporised position, DEMCA framework can handle more realistic contracts deadlines. 
Another difference consists in the contract representation, in our case the contract is an extension of a social commitment in accordance with practice in law: gratuitous promise, unilateral contract, bilateral contract. 
By introducing defeasible logic when reasoning about commitments, we obtain two main advantages. On the one hand, agents can reason with incomplete information. Also, this property of
nonmonotonic logics permits, in our case, to model confidential
contractual clauses. 
Expected exceptions can be captured by defining a preference structure 
over the runs within the commitments dependency network~\cite{mallya05aamas}.
Having the superiority relation from defeasible logic, 
we can easily define such a structure in our framework.

\section{CONCLUSIONS}
\label{sec:conclusions}

This paper introduces the DEMCA framework for designing electronic markets for contractual agents. 
While the amount of existing literature on e-commerce focuses on contract negation or contract formation, we focus on the contract breach phase during the lifecycle of a contract.

The contribution contains two ideas. On the one hand, we apply the
principles of contract law in the task dependency network
model~\cite{Walsh03}. As a result, we enrich that model by
including different types of penalties when agents breach, thus
bringing the model closer to the real world.
On the other hand, the implemented framework can be used for simulations when designed electronic markets, knowing that the dispute resolution mechanism is a key factor in the success of e-commerce applications~\cite{Rule06AIL}. 
Such a framework is useful for automated online dispute resolution. The data obtained can be used as arguments in a mediated dispute or the remedies can be computed in real time
in case the agents agreed with the market policy.

\bibliographystyle{plain}
\bibliography{denca}

\end{document}